\documentclass{acm_proc_article-sp}
\usepackage{color}
\usepackage{multirow}
\usepackage{graphicx}
 \usepackage{soul}

\newcommand{\comment}[1]{}
\definecolor{Orange}{rgb}{1,0.5,0}

\begin{document}

\title{Stacking from Tags: \\ Clustering Bookmarks around a Theme}

\numberofauthors{2}

\author{}
\author{
\alignauthor Arkaitz Zubiaga\\
       \affaddr{Queens College and Graduate Center}\\
       \affaddr{City University of New York}\\
       \affaddr{New York, NY, USA}\\
       \email{arkaitz@zubiaga.org}
\alignauthor Alberto~P\'{e}rez~Garc\'{i}a-Plaza, V\'{i}ctor~Fresno, Raquel~Mart\'{i}nez\\
       \affaddr{NLP \& IR Group}\\
       \affaddr{ETSI Inform\'{a}tica UNED}\\
       \affaddr{Madrid, Spain}\\
       \email{\{alpgarcia, vfresno, raquel\}@lsi.uned.es}
}

\maketitle
\begin{abstract}
 Since very recently, users on the social bookmarking service Delicious can stack web pages
in addition to tagging them. Stacking enables users to group web pages around specific themes with
the aim of recommending to others. However, users still stack a small subset of what they tag, and
thus many web pages remain unstacked. This paper presents early research towards automatically
clustering web pages from tags to find stacks and extend recommendations.
\end{abstract}

\section{Introduction}

Social tagging has become a powerful means to organize resources facilitating later
access \cite{golder2006usage}. On these sites, users can label web pages with tags, facilitating future
access to those web pages both to the author and to other interested users \cite{heymann2008can,zubiaga2009getting}.
Very recently, Delicious.com introduced a new dimension for organizing web pages:
stacking. With stacks, users can group the web pages that might be of interest for a specific
community, e.g., \textit{Valentine's Special}, or \textit{UNIX and Programming Jokes}. Stacks can be
very useful for those who are looking for help or information on a specific matter. With stacks,
users are providing a 2-dimensional organization of web pages that is complemented with tags, as
shown by the example in Table \ref{tab:tagging-stacking}. However, tagging activity still clearly
exceeds the stacking activity, and many web pages are tagged but not stacked. Moreover, all the web
pages tagged before the new feature was introduced are not associated with stacks. Thus, finding a way
to infer missing stacks from all those tags would be helpful to recommend more groups of web
pages to communities, or to suggest adding web pages to any user stack. Different from previous
research on clustering and classifying tagged resources, which evaluated using the Open Directory Project \cite{ramage2009clustering,zubiaga2012harnessing} or from manually built categorizations \cite{zubiaga2009content}, stacks provide a rather ad hoc ground truth to
evaluate with. This paper describes early research in this direction, presenting preliminary work on
automatically clustering web pages from tags to find stacks as users would do. Preliminary
experiments suggest that using tags can help reach high performance clustering.

\begin{table}
 \begin{center}
  \scriptsize
  \begin{tabular}{ c  c  c  c  c  c  c }
   \multicolumn{3}{c}{Stack \#1} & \multicolumn{1}{c}{} & \multicolumn{2}{c}{Stack \#2} &
\multicolumn{1}{c}{} \\
   \multicolumn{3}{c}{\rule[1pt]{80pt}{2pt}} & \multicolumn{1}{c}{} & \multicolumn{2}{c}{\rule[1pt]{50pt}{2pt}} & \multicolumn{1}{c}{} \\
   URL$_1$ & URL$_2$ & URL$_3$ & URL$_4$ & URL$_5$ & URL$_6$ & URL$_7$ \\
    \multicolumn{1}{c}{\rule[1pt]{20pt}{1pt}} & \multicolumn{1}{c}{\rule[1pt]{20pt}{1pt}} &
    \multicolumn{1}{c}{\rule[1pt]{20pt}{1pt}} & \multicolumn{1}{c}{\rule[1pt]{20pt}{1pt}} &
    \multicolumn{1}{c}{\rule[1pt]{20pt}{1pt}} & \multicolumn{1}{c}{\rule[1pt]{20pt}{1pt}} &
    \multicolumn{1}{c}{\rule[1pt]{20pt}{1pt}} \\
   tags$_1$ & tags$_2$ & tags$_3$ & tags$_4$ & tags$_5$ & tags$_6$ & tags$_7$ \\
  \end{tabular}
  \caption{Example of a user's tags and stacks. The user tagged 7 URLs, with 5 of them in 2 stacks.}
  \label{tab:tagging-stacking}
 \end{center}
\end{table}

\section{Dataset}

We collected the tagging activity for 3,635 Delicious users in October and November 2011. This
subset includes all the users who created at least a stack in this timeframe. During this period,
those users tagged 182,510 web pages, creating 5,214 stacks. Out of those web pages, 45,196 (24.8\%)
were stacked while 137,314 (75.2\%) were left out of stacks. Also, a large set of users who are
not included in our dataset are tagging web pages, while they are not creating stacks. Going into
further details on the tagging activity in and out of the stacks, we observe that, on average,
30.1\% of the tags contained in stacks are also used out of the stacks. This suggests that there is
not specific vocabulary for stacks, but users share vocabulary with web pages out of stacks.
Moreover, there is just a small subset of 22.5\% of the stacks that have a common tag in all their
underlying web pages. Hence, most users do not use an exclusive tag that refers to the stack. This
motivates our study on the automatic clustering of web pages from tags with the aim of finding
stacks that approach to those created by users.

\section{Experiments}
We used \textit{Cluto rbr} \cite{karypis03} (k-way 
repeated bisections globally optimized) to find clusters from tags. 
\textit{Cluto rbr} conveniently fits with the present task since, in practice, it always generates
the same clustering solution for a certain input data. As the main parameter,
this algorithm requires as an input the number of clusters to generate, which is known as
\textit{K}. We used values ranging from 2 to 10 for \textit{K}, as a preliminary approach that
allows us to evaluate and understand how the number of created clusters affects the solution. We set
the rest of the parameters to their default values. Upon these settings, we got the resulting
clusters for all the web pages saved by each user, and compare the results to the stack(s) created
by the user. For each run on a stack, we computed the precision, recall and F1 values, and got the
macroaveraged values for all the stacks.

\begin{figure}[tbh]
 \begin{center}
  \includegraphics[width=200px]{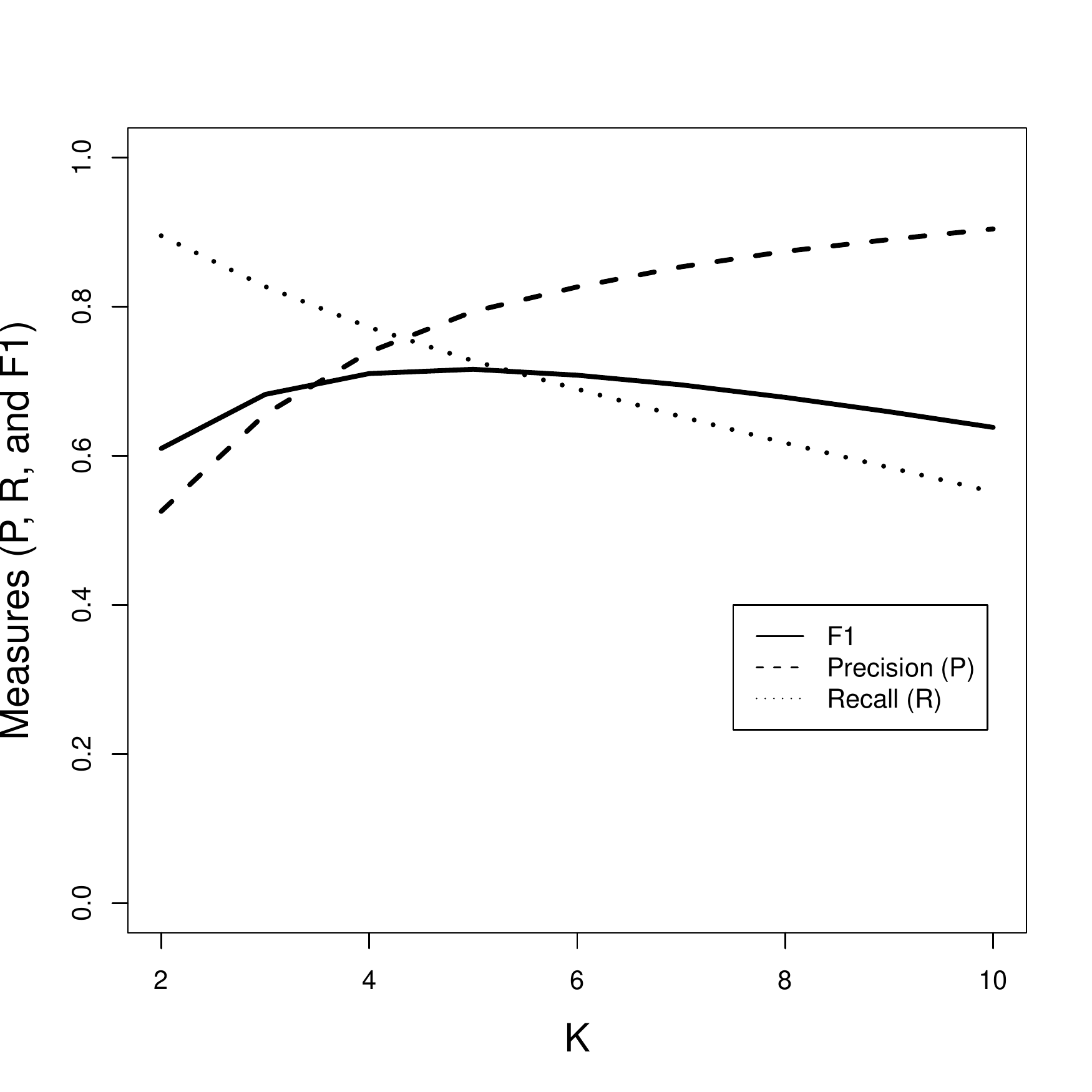}
  \caption{Macroaveraged F1, P and R for K-dependent Cluto runs, ranging from 2 to 10 for K.}
  \label{fig:results-own}
 \end{center}
\end{figure}

Figure \ref{fig:results-own} shows precision, recall, and F1 values for K-dependent Cluto runs.
Precision and recall values considerably vary depending on the selected \textit{K} value.
Creating a few big clusters improves recall, while creating many small clusters improves precision.
However, F1 values remain very similar while \textit{K} changes. Regardless of the value selected
for \textit{K}, the clustering gets F1 above 0.6. Hence, the selection of the value
for \textit{K} mainly conditions that the results get affected by either precision or recall,
depending on the preference.

\begin{figure}[tbh]
 \begin{center}
  \includegraphics[width=200px]{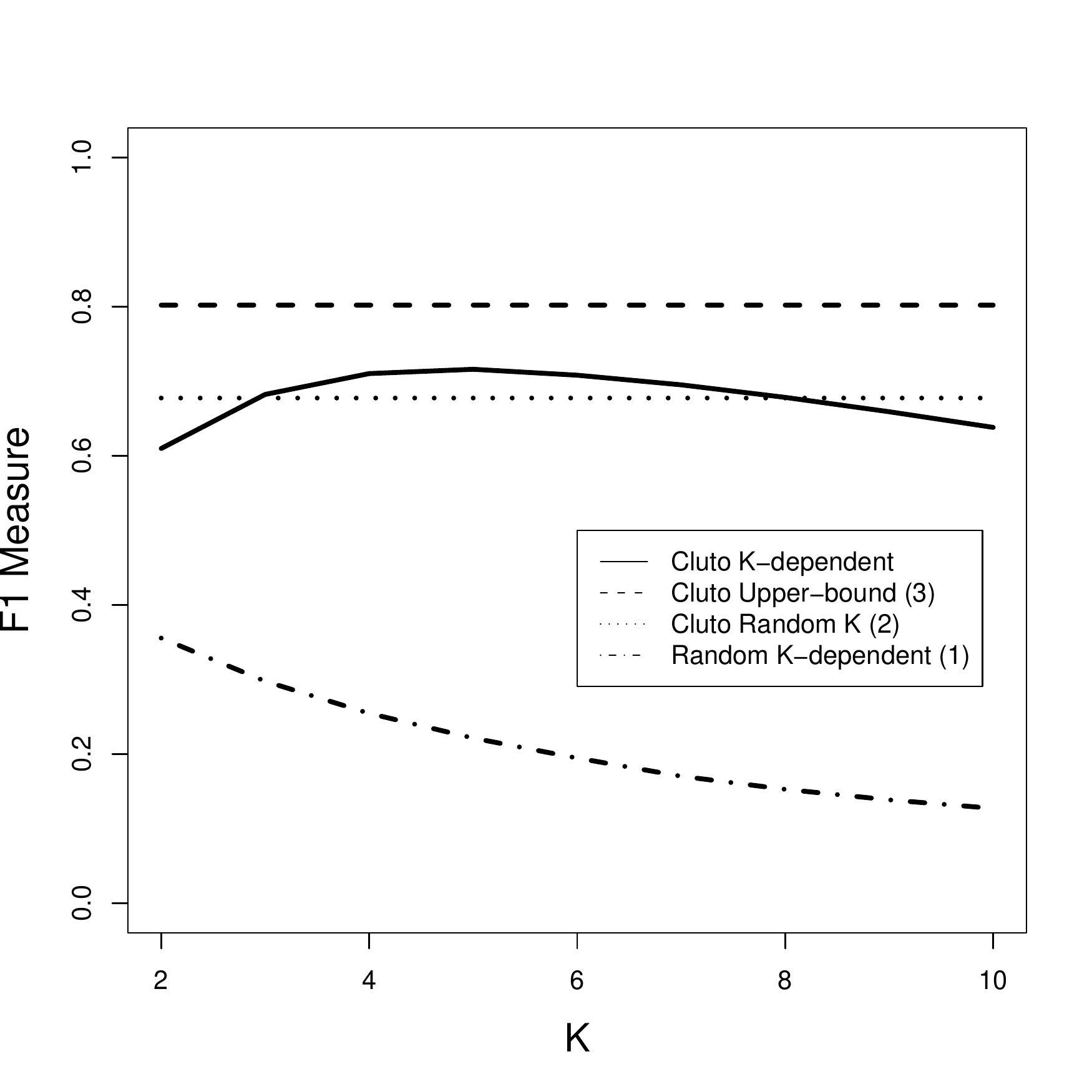}
  \caption{Comparison of F1 values for K-dependent Cluto runs as compared to benchmark approaches.}
  \label{fig:results-compare}
 \end{center}
\end{figure}

Figure \ref{fig:results-compare} complements the above results by showing the F1 values for our
K-dependent runs
as compared to 3 benchmark methods: (1) a baseline approach that randomly creates the clusters,
i.e., randomly generating \textit{K} clusters of equal size, (2) an intermediate approach that
randomly selects the value of \textit{K} for each user, i.e., the average of multiple runs
using random \textit{K} values, and (3) the ideal upper-bound performance by choosing the optimal
\textit{K} value for each user stack. These results show that using tags to find stacks clearly
outperforms a random approach, doubling the performance in many cases. This encourages the use of
tags to perform this task in an effective way. Moreover, even though this paper does not explore
how to find a suitable \textit{K} for each stack, the upper-bound performance based on optimal
\textit{K} values shows that tags can reach very high results. An appropriate selection of
\textit{K} could yield clusters approaching to 0.8 performance in terms of F1. It also clearly
outperforms the random selection of \textit{K}, encouraging to perform further research in a
way of looking for a suitable \textit{K} for each user.

\section{Conclusions}

This work describes early research for a work-in-progress on a novel feature of social bookmarking
systems: stacking. To the best of our knowledge, this is the first research work that deals with
stacks. We have shown that the use of tags to find stacks that resemble to those created by users
scores high performance results above 0.6 in terms of F1. Moreover, choosing the right parameters
for each stack to be created can substantially improve performance by scoring nearly 0.8. As a
preliminary work, these results encourage performing further study that helps make a decision on
the selection of parameters that improves performance. Future work includes studying
behavioral patterns of users such as tagging vocabulary towards finding the right parameters for
each user. The promising results by using tags to discover stacks also suggest further research
looking for groups of related tags both to individual users and communities.

\bibliographystyle{abbrv}
\bibliography{sigir2012-stacks}

\end{document}